\documentclass[pra,aps,twocolumn,showpacs,nofootinbib,longbibliography]{revtex4-1}

\usepackage{placeins}  
\usepackage{amsfonts, amsmath,amssymb,amsthm,bm}
\usepackage{mathrsfs}
\usepackage{graphicx}
\usepackage[usenames,dvipsnames]{color}
\usepackage{palatino}
\usepackage{fancyhdr}
\usepackage{mathrsfs}
\usepackage{multirow}
\usepackage{hyperref}

\hypersetup{
   colorlinks=true,         
   linkcolor=blue,          
   citecolor=blue,          
   urlcolor=Violet          
}

\newcommand{\bra}[1]{\ensuremath{\left\langle#1\right|}}
\newcommand{\ket}[1]{\ensuremath{\left|#1\right\rangle}}

\newcommand{\ketbra}[2]{\ensuremath{\left|#1\right\rangle\!\left\langle#2\right|}}

\newcommand{\tr}[2]{\mathrm{Tr}_{#1}\left( #2 \right)}

\renewcommand{\v}[1]{\ensuremath{\boldsymbol #1}}

\newcommand{\be}{\begin{equation}}
\newcommand{\ee}{\end{equation}}

\newcommand{\E}{\mathcal{E}}

\newcommand{\I}{\mathbb{I}}

\theoremstyle{plain}

\theoremstyle{definition}

\theoremstyle{remark}

\begin{document}

\title{Quantum coherence, time-translation symmetry and thermodynamics}
\author{Matteo Lostaglio*, Kamil Korzekwa*, David Jennings and Terry Rudolph}
\affiliation{Department of Physics, Imperial College London, London SW7 2AZ, United Kingdom}
\footnotetext{These authors contributed equally to this work.}
\begin{abstract}
The first law of thermodynamics imposes not just a constraint on the energy-content of systems in extreme quantum regimes, but also symmetry-constraints related to the thermodynamic processing of quantum coherence. We show that this thermodynamic symmetry decomposes any quantum state into mode operators that quantify the coherence present in the state. We then establish general upper and lower bounds for the evolution of quantum coherence under arbitrary thermal operations, valid for any temperature. We identify primitive coherence manipulations and show that the transfer of coherence between energy levels manifests irreversibility not captured by free energy. Moreover, the recently developed thermo-majorization relations on block-diagonal quantum states are observed to be special cases of this symmetry analysis.
\end{abstract}

\pacs{03.65.Ta, 03.67.-a, 05.70.Ln}
\maketitle

\section{Introduction}

Fundamental laws of Nature often take the form of restrictions: nothing can move faster than light in vacuum, energy cannot be created from nothing, there are no perpetuum mobiles. It is due to these limitations that we can ascribe value to different objects and phenomena, e.g., energy would not be treated as a resource if we could create it for free. The mathematical framework developed to study the influence of such constraints on the possible evolution of physical systems is known under the collective name of resource theories.

Perhaps the best known example of this approach was to formalize and harness the puzzling phenomenon of quantum entanglement (see \cite{horodecki2009quantum} and references therein). However, the basic machinery developed to study entanglement is also perfectly suited to shed light on a much older subject -- thermodynamics. The first and second laws are fundamental constraints in thermodynamics. These force thermodynamic processes to conserve the overall energy and forbid free conversion of thermal energy into work. Thus, a natural question to ask is: what amounts to a resource when we are restricted by these laws? This question is particularly interesting in the context of small quantum systems in the emergent field of single-shot thermodynamics \cite{janzing2000thermodynamic, brandao2011resource, horodecki2013fundamental, brandao2013second, aberg2013truly, skrzypczyk2013extracting, aberg2014catalytic, halpern2014unification, dahlsten2011inadequacy}.

\emph{Athermality} is the property of a state of having a distribution over energy levels that is not thermal \cite{brandao2011resource}. This is a resource because, as expected from the Szilard argument \cite{szilard1929uber}, it can be converted into work \cite{egloff2012laws, brandao2013second}, which in turn can be used to drive another system out of equilibrium. However \emph{coherence} can be viewed as a second, independent resource in thermodynamics \cite{lostaglio2015description}. This stems from the fact that energy conservation, implied by the first law, restricts the thermodynamic processing of coherence. Hence possessing a state with coherence allows otherwise impossible transformations. Energy conservation also enforces a modification of the traditional Szilard argument: both athermality and coherence contribute to the free energy, however coherence remains ``locked'' and cannot be extracted as work~\cite{skrzypczyk2013extracting,lostaglio2015description}.

Since coherence is a thermodynamic resource, an open question is what kind of coherence processing is allowed by thermodynamic means. This foundational question is of interest for future advancements in nanotechnology, as interference effects are particularly relevant \cite{karlstrom2011increasing, vazquez2012probing} at scales we are increasingly able to control \cite{collin2005verification, serreli2006molecular, toyabe2010experimental,alemany2010fluctuations, cheng2010bipedal}. Moreover, recent evidence suggests that biological systems may harness quantum coherence in relevant timescales \cite{lloyd2011quantum,lambert2013quantum,gauger2011sustained}. Despite partial results \cite{scully2003extracting, rodriguez2013thermodynamics, skrzypczyk2013extracting, aberg2014catalytic, cwiklinski2014limitations, narasimhachar2014low} we still lack a complete understanding of the possible coherence manipulations in thermodynamics. The aim of this paper is to address this problem making use of recently developed tools from the resource theory of asymmetry \cite{marvian2013modes, marvian2014extending}.


\section{The paradigmatic setting}

The central question of thermodynamics is: what are the allowed transformations of a system that are consistent with the first and second laws? Much of the developments in single-shot thermodynamics have been restricted to quantum states that do not possess quantum coherence between energy eigenspaces \cite{egloff2012laws, horodecki2013fundamental, brandao2013second, aberg2013truly}, and the recent analysis has shown that a whole family of independent entropic measures provide necessary \emph{and sufficient} conditions when the states are incoherent in energy (free energies $F_\alpha$, parametrized by a real number $\alpha$, must all decrease \cite{brandao2013second}). However it was established in \cite{lostaglio2015description} that quantum coherence \emph{cannot} be properly described by free energy relations, and that new and independent relations are required. These new constraints originate from energy conservation in thermodynamics and were derived from the resource theory that quantifies the degree to which a quantum state lacks time-translation invariance.

\begin{figure}[t!]
\includegraphics[width=0.6\columnwidth]{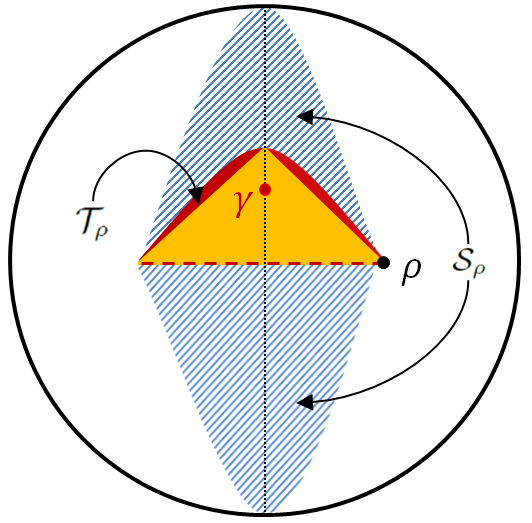}
\caption{\emph{The basic structure}. \label{fig:achievable} The set of states $ \mathcal{T}_\rho \,$ achievable under thermal operations from the initial qubit state $\rho$ is given by the dark red solid region and the orange triangle ($\gamma$ is the thermal state of the system). If thermal operations on coherent states were trivial, in the sense that they were equivalent to dephasings and operations on incoherent states, then this set would reduce to the orange triangle. Moreover, even if one has access to arbitrary amount of work (but not coherence), then the set of achievable state is extended to the dashed blue region $ \mathcal{S}_\rho \,$, but not to the whole Bloch sphere.}
\end{figure}

Let us set the scene with a transparent example that illustrates the issues at hand. The simplest possible example is a qubit system with Hamiltonian $H_S= \ketbra{1}{1}$ that can interact with arbitrary heat baths at temperature $kT=\beta^{-1}$, through energy-conserving interactions on the composite system (these maps are called \emph{thermal operations}, see Sec.~\ref{sec:thermal}). The thermal state of the system is given by $\gamma = e^{-\beta H_S}/\tr{}{e^{-\beta H_S}}$. The core question now is: given a qubit state $\rho$ that possesses quantum coherence, what is the set of quantum states $ \mathcal{T}_\rho \,$ accessible from $\rho$ under thermal operations? Its basic structure is that of a rotationally symmetric (about the Z-axis), convex set of states. In the X-Z plane of the Bloch sphere, this set is given by the dark red solid region and the orange triangle, see Fig.~\ref{fig:achievable}. The boundary surface denotes the states that preserve the maximal amount of coherence while having a given final energy distribution. Let us analyse the structure of $ \mathcal{T}_\rho \,$ in a more detailed way to show the non-triviality of coherence transformations in thermodynamics.

First of all one might expect that, due to the intrinsically dissipative interactions of a quantum system with the heat bath, coherence is only playing a passive role in the process. If this was the case all possible transformations would be attainable just by a combination of dephasings and thermal operations on incoherent states. However, the set of states achievable in this way is limited to the orange triangle in Fig.~\ref{fig:achievable} and clearly does not coincide with $ \mathcal{T}_\rho \,$ (details will be given later). We conclude that coherence is actively contributing to enlarge the set of thermodynamically accessible states.  

One might also ask the following question: if we are given an unbounded amount of free energy, would coherence still be a resource? If the answer was no, all constraints could be lifted by a sufficiently large work source. Work would be \emph{the} universal resource of thermodynamics. However, this is not the case. To see this suppose the unbounded amount of work is given in the form of arbitrary number of copies of pure, zero-coherence states, say $\ket{0}^{\otimes N}$. It is easy to show (see Appendix B) that allowing arbitrary amounts of work only extends $ \mathcal{T}_\rho \,$ to the set of states $\mathcal{S}_{\rho}$ accessible under ``time-symmetric evolutions'' (dashed blue region in Fig.~\ref{fig:achievable}), which is a strict subset of the full Bloch sphere. Therefore, we can conclude that work is not a universal resource and coherence resources should be carefully accounted for.

Finally, this analysis shows that the classical Szilard argument linking classical information and thermodynamics does not simply carry over to the quantum regime. Classically we know that a single bit has an ``energetic value'' of $kT\ln 2$, and so one might also expect that the possession of a single pure qubit state allows for extracting $kT \ln 2$ of mechanical work from a heat bath. However, consider the qubit state \mbox{$\ket{\psi_\beta} \propto \ket{0}+\sqrt{e^{-\beta}} \ket{1}$}. The analysis of \cite{skrzypczyk2013extracting, lostaglio2015description} shows that, due to total energy conservation implied by the first law of thermodynamics, it is fundamentally impossible to distinguish this state from the Gibbs equilbrium state $\gamma$. Therefore no work can be extracted from such pure state unless we have access to \emph{coherence resources} \cite{aberg2014catalytic}.


\section{Thermal Operations and Symmetries}
\label{sec:thermal}

\subsection{Thermal operations}

We consider the following general setting for thermodynamic transformations. A quantum system, previously isolated and characterized by a Hamiltonian $H_S$, is brought into thermal contact with a bath described by a Hamiltonian $H_E$. After some time the system is decoupled from the bath. The only assumption we make is that this interaction conserves energy \emph{overall} (of course heat will flow from and to the bath), according to the first law of thermodynamics. Mathematically this can be formalised through the notion of \emph{thermal operations} \cite{janzing2000thermodynamic, brandao2011resource, horodecki2013fundamental, brandao2013second}, i.e., the set of maps $\{\mathcal{E}_T\}$ that act on a system $\rho$ in the following way:
\begin{equation}
\label{eq:thermal}
\mathcal{E}_T(\rho)=\tr{E}{U\left[\rho\otimes \gamma_E\right]U^{\dagger}},
\end{equation}
where $U$ is a joint unitary commuting with the total Hamiltonian of the system and environment, \mbox{$[U, H_S + H_E]=0$}, and $\gamma_E$ is a thermal (Gibbsian) state of the environment at some fixed inverse temperature $\beta$, \mbox{$\gamma_E = e^{-\beta H_E}/\tr{}{e^{-\beta H_E}}$}. 

As observed in \cite{lostaglio2015description} we identify two main properties of thermal maps:
\begin{enumerate}
\item $\{\mathcal{E}_T\}$ are time-translation symmetric \cite{marvian2014extending}, i.e.,  
\be
\label{eq:symmetric}
\mathcal{E}\left(e^{-i H_S t} \rho e^{i H_S t}\right) = e^{-i H_S t} \mathcal{E}(\rho) e^{i H_S t}.
\ee
\item $\{\mathcal{E}_T\}$ preserve the Gibbs state,
\be
\label{eq:gp}
\mathcal{E}_T(\gamma)=\gamma.
\ee
\end{enumerate}
The first property reflects energy conservation, a consequence of the first law, and the fact that the thermal bath is an incoherent mixture of energy states. The second property incorporates the core physical principle of the second law of thermodynamics: the non-existence of a machine able to run a cycle in which thermal energy is converted into work. Eq.~\eqref{eq:gp} requires that we cannot bring a thermal state out of equilibrium at no work cost. Indeed, if this was the case we could equilibrate it back and extract work, giving us a perpetuum mobile of the second kind. 

\subsection{Modes of coherence}

In the current work we use the fact that symmetric operations, i.e. maps satisfying \eqref{eq:symmetric}, naturally decompose quantum states into ``modes''. Modes can be seen as a generalization of Fourier analysis to the context of operators \cite{marvian2013modes}. Physically, they identify components within a quantum state that transform independently as a consequence of the underlying symmetry of the dynamics.

The theory introduced in \cite{marvian2013modes} can be easily adapted to thermodynamics. Let us expand the system state $\rho$ in the eigenbasis of its Hamiltonian $H_S$ as follows:
\begin{equation*}
\rho = \sum_{n,m} \rho_{nm} \ketbra{n}{m},
\end{equation*}
where \mbox{$H_S \ket{n} = \hbar\omega_n \ket{n}$}. We limit our considerations here to non-degenerate $H_S$, as thermal operations allow for any energy preserving unitary to be performed on the system. This means that there are no limitations on transferring coherence between different degenerate energy levels, which gives rise to additional structure within each degenerate energy subspace obscuring the general picture. Let us now denote the set of all differences between eigenfrequencies of $H_S$ by $\{\omega \}$ . Then
\be
\label{eq:modes}
\rho = \sum_{\omega} \rho^{(\omega)}, \quad \rho^{(\omega)} := \sum_{\substack{n,m \\ \omega_n - \omega_m = \omega}} \rho_{nm} \ketbra{n}{m}.
\ee
The operators $\rho^{(\omega)}$ are \emph{modes of coherence} of the state $\rho$. Modes are characterized by their transformation property under the symmetry group:
\be
e^{-i H_S t} \rho^{(\omega)} e^{i H_S t} = e^{-i \omega t} \rho^{(\omega)},
\ee
and are therefore 1-dimensional irreps of the U(1) time-translation group action. It is easy to check that if $\mathcal{E}_T$ is a thermal operation (so also symmetric) such that \mbox{$\mathcal{E}_T(\rho)=\sigma$}, then 
\be
\label{eq:covariance}
\mathcal{E}_T(\rho^{(\omega)}) = \sigma^{(\omega)} \quad \quad \forall \omega.
\ee
In other words each mode $\rho^{(\omega)}$ in the initial state is independently mapped by a thermal operation to the corresponding mode $\sigma^{(\omega)}$ of the final state.

Eq.~\eqref{eq:covariance} allows us to introduce natural measures of coherence for each mode, as shown in \cite{marvian2013modes}. Since the $1$-norm is contractive under general quantum operations, we have for any bounded linear operator $X$,
\be
\label{eq:contractive}
||\mathcal{E}(X)||_1 \leq ||X||_1, \quad ||X||_1 := \tr{}{\sqrt{X X^{\dag}}}.
\ee
Now, Eqs.~\eqref{eq:contractive} and \eqref{eq:covariance} together imply that the total amount of coherence in each mode is non-increasing under thermal operations. For all $\omega$,
\begin{equation}
\label{eq:monotone}
\sum_{\substack{n,m \\ \omega_n - \omega_m = \omega}} |\sigma_{nm}| \leq\sum_{\substack{n,m \\ \omega_n - \omega_m = \omega}} |\rho_{nm}|.
\end{equation}
Note that these constraints are only due to the symmetry properties of thermal operations, and therefore would hold in a situation where we allow arbitrary amounts of work to be available, as previously discussed. 

\subsection{Thermomajorization as a zero mode constraint}

Necessary and sufficient conditions for thermodynamic interconversion between states block-diagonal in the energy eigenbasis have been recently found \cite{horodecki2013fundamental}. Given an initial incoherent state $\rho = \rho^{(0)}$, a final state $\sigma = \sigma^{(0)}$ is thermodynamically accessible if and only if 
\be
\label{eq:zeromode}
\sigma^{(0)} \prec_T \rho^{(0)},
\ee
where $\prec_T$ is a generalisation of majorization \cite{ruch1978mixing,marshall2010inequalities}, called thermomajorization. Eq.~\eqref{eq:covariance} shows that given two general quantum states $\rho$ and $\sigma$, for $\sigma$ to be thermally accessible from $\rho$, a set of independent equations must be simultaneously fulfilled. The thermomajorization condition of Eq.~\eqref{eq:zeromode} only ensures that Eq.~\eqref{eq:covariance} is satisfied for the $\omega=0$ mode, leaving open the question of the thermodynamic constraints on coherent transformations on all nonzero modes.

\section{Bounds on coherence transformations}
\label{sec:bounds}

The conceptual framework just described provides a natural way to analyse coherence within thermodynamics. We now develop both upper and lower bounds on how the modes of coherence evolve under general thermodynamic transformations. 

\subsection{Lower bound on guaranteed coherence preservation}

Consider an initial state $\rho$, with energy measurement statistics given by $\rho^{(0)}$ (by this, we mean the spectrum of $\rho^{(0)}$). Suppose we want to modify this distribution into a new distribution $\sigma^{(0)}$. From Eq.~\eqref{eq:zeromode} this is possible if and only if $\sigma^{(0)} \prec_T \rho^{(0)} $. The question is now: how much quantum coherence can be preserved in this process? Here we will establish a lower bound on the guaranteed coherence for such a transformation that relies only on known results about thermodynamic transformations among incoherent states \cite{horodecki2013fundamental} and the convexity of the set of thermal operations. 

Assume that there exists a thermal operation mapping $\rho^{(0)}$ into $\sigma^{(0)}$. Define $\Sigma$ as the set of quantum states with a distribution over the energy eigenstates given by $\sigma^{(0)}$ and denote by $\mathcal{T}_\rho$ the set of states accessible from $\rho$ through thermal maps. It is easy to see that \mbox{$\Sigma \cap \mathcal{T}_\rho \neq\emptyset$}, because the dephasing operation $\rho \mapsto \rho^{(0)}$ is a thermal operation. 
Hence, it is natural to ask which state in this intersection has the highest amount of coherence? 

\begin{figure}[t!]
\includegraphics[width=0.7\columnwidth]{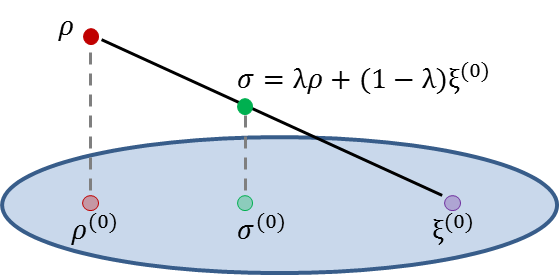}
\caption{\label{fig:guaranteed}\emph{Guaranteed coherence.} The shaded region represents the set of incoherent states. By convexity of the set of thermal operations, if $\xi^{(0)}$ is thermally achievable from $\rho^{(0)}$, then also $\sigma$ can be achieved from $\rho$.}
\end{figure}

First consider the set $\mathcal{T}_{\rho^{(0)}}$, which is contained in $\mathcal{T}_\rho$ and is completely characterized by thermomajorization \cite{horodecki2013fundamental}. Within this set, consider the states \mbox{$\{ \xi=\xi^{(0)}: \sigma^{(0)} = \lambda \rho^{(0)} + (1-\lambda) \xi^{(0)}\}$} along the line of $\rho^{(0)}$ and $\sigma^{(0)}$. From any of these we can define a state \mbox{$\sigma = \lambda \rho + (1-\lambda) \xi^{(0)}$} (see Fig.~\ref{fig:guaranteed}). One can check that $\sigma \in \Sigma$ and that $\sigma$ is a convex combination of two states in $\mathcal{T}_\rho$ ($\rho \in \mathcal{T}_\rho$ trivially and by definition $\xi^{(0)} \in \mathcal{T}_{\rho^{(0)}} \subseteq \mathcal{T}_\rho$). Moreover, we can show that the set of thermal maps is a convex set (see Appendix C), and so $\mathcal{T}_\rho$ is also convex. This immediately implies $\sigma\in\mathcal{T}_{\rho}$. 

Now note that the modes $\sigma^{(\omega)}$ of the final state $\sigma$ (as defined in Eq.~\eqref{eq:modes}) can only come from the initial state $\rho$, as $\xi^{(0)}$ has zero coherence. Therefore, we conclude that the fraction $\lambda$ gives a lower bound on the coherence that can be preserved in each mode, as \mbox{$\sigma^{(\omega)}= \lambda \rho^{(\omega)}$}. By extremizing $\xi^{(0)}$ within the set $\mathcal{T}_{\rho^{(0)}}$ we obtain the optimal fraction $\lambda=\lambda_{*}$ of guaranteed coherence in each mode:
\be
\label{eq:guaranteed}
\sigma^{(\omega)}= \lambda_{*} \rho^{(\omega)}.
\ee
As we show in Appendix G (specifically see Fig.~\ref{fig:qubit_cov_th}) this lower bound is not tight already in the simplest scenario of a qubit system. This indicates that there is more to the thermal inter-conversion of quantum states than simply a combination of dephasing and thermodynamic transformations on incoherent states.

\subsection{Maximal coherence}

We will now derive an upper bound on the coherence in the final state dependent on the transition probabilities between energy levels. Consider the open quantum system dynamics described by unitarily coupling a system $\rho$ with an initially uncorrelated environment $E$ in state $\tau$:
\be
\label{eq:cptp}
\mathcal{E}(\rho)= \tr{E}{U(\rho \otimes \tau)U^{\dag}}.
\ee
By the Stinespring dilation theorem, any completely positive trace-preserving (CPTP) map can be realized in this way \cite{nielsen2010quantum}. Expanding $\tau$ in its eigenbasis as $\tau = \sum_a \lambda_a \ketbra{a}{a}$, every map \eqref{eq:cptp} can be rewritten as \cite{breuer2002open}
\be
\label{eq:wcptp}
\mathcal{E}(\rho) = \sum_{a,b} W_{ab} \rho W^{\dag}_{ab},
\ee
where $W_{ab} = \sqrt{\lambda_a} \bra{b}U\ket{a}$. The final off-diagonal element (coherence between energy states) \mbox{$\rho'_{nm} = \bra{n}\mathcal{E}(\rho)\ket{m}$} can be written as
\begin{equation*}
\rho'_{nm} = \sum_{c,d} \rho_{cd} \sum_{a,b} \bra{n}W_{ab}\ket{c}\bra{d} W^{\dag}_{ab}\ket{m}.
\end{equation*}
Defining the matrix $X^{(xy)}$ whose elements are \mbox{$X^{(xy)}_{ab} = \bra{y}W_{ab}\ket{x}$} we obtain
\begin{equation*}
\rho'_{nm} = \sum_{c,d} \rho_{cd} \tr{}{X^{(cn)} X^{(dm) \dag}}.
\end{equation*}
Obviously $|\rho'_{nm}| \leq \sum_{cd} |\rho_{cd}| |\tr{}{X^{(cn)} X^{(dm)\dag}}|$.
Using the Cauchy-Schwarz inequality,
\small
\be
\label{eq:cauchy}
|\rho'_{nm}| \leq \sum_{c,d} |\rho_{cd}| \sqrt{\tr{}{X^{(cn)}  X^{(cn) \dag}}\tr{}{X^{(dm)}  X^{(dm) \dag}}}.
\ee
\normalsize
We can define $p_{n|c}$ as elements of the stochastic matrix $\Lambda$ induced on the diagonal elements of the quantum state,
\be
\label{eq:stochastic}
p_{n|c} = \bra{n}\mathcal{E}(\ketbra{c}{c})\ket{n}.
\ee
The matrix $\Lambda$ is stochastic, because $\mathcal{E}$ is trace-preserving. Inserting \eqref{eq:wcptp} in \eqref{eq:stochastic} we can check that
\begin{equation*}
p_{n|c}=\sum_{a,b} |\bra{n}W_{ab}\ket{c}|^2 = \tr{}{X^{(cn)}  X^{(cn)\dag}},
\end{equation*}
so that substituting the above into Eq. \eqref{eq:cauchy}, we arrive at a bound for processing coherence under a general CPTP map: 
\be
\label{eq:cptpbound}
|\rho'_{nm}| \leq \sum_{c,d} |\rho_{cd}| \sqrt{p_{n|c} p_{m|d}}.
\ee

\subsubsection*{Time-translation symmetry condition}

The fact that thermal operations \eqref{eq:thermal} are symmetric greatly simplifies the bound \eqref{eq:cptpbound}. From the property \eqref{eq:covariance} each mode of a quantum state transforms independently. This immediately implies that we can refine \eqref{eq:cptpbound} to get 
\be
\label{eq:covariantbound}
|\rho'_{nm}| \leq \sideset{}{'}\sum_{c,d} |\rho_{cd}| \sqrt{p_{n|c} p_{m|d}},
\ee 
where the primed sum $\sum'$ denotes the summation only over indices $c$ and $d$ such that $\omega_c-\omega_d = \omega_n-\omega_m$. Thus the given final coherence between states differing by $\hbar\omega$ in energy can only come from initial coherences between pairs of states that differ in energy also by $\hbar\omega$. We note that the recent result of \cite{cwiklinski2014limitations} is a special case of the above bound where no summation occurs. The broad structure of the bound given by Eq.~\eqref{eq:cptpbound} holds for any CPTP map and the result of \cite{cwiklinski2014limitations} simply encodes energy conservation in the restricted case of no splitting degeneracies. Finally, let us emphasize that the bound \eqref{eq:covariantbound} applies not only to thermal operations, but more generally to all time-translation symmetric maps, i.e., all quantum operations satisfying \eqref{eq:symmetric}. Further thermodynamic constraints purely due to symmetry are analyzed in \cite{lostaglio2015description}.

\subsubsection*{Gibbs-preserving condition}

The bound \eqref{eq:covariantbound} can be refined further by noting that the Gibbs-preserving condition~\eqref{eq:gp} puts restrictions on the transition probabilities $p_{l|k}$. Specifically, it induces the following equality:
\begin{equation}
\label{eq:GP_cond}
\Lambda \v{r}=\v{r},
\end{equation}
where $\v{r}=(r_0\dots r_{d-1})$ denotes the vector of thermal probabilities of the $d$-dimensional system under consideration and $\Lambda$ is the matrix whose elements $p_{l|k}$ are defined by Eq.~\eqref{eq:stochastic}. From \eqref{eq:GP_cond} one can prove that (see Appendix~D for details):
\begin{equation}
\label{eq:transitionbound}
p_{l|k}\leq e^{\beta\hbar(\omega_k-\omega_l)}\quad \forall_{k,l}.
\end{equation}
Hence, if the energy of the final state $\hbar\omega_l$ is higher than the energy of the initial state $\hbar\omega_k$, the transition probability is bounded by $e^{-\beta\hbar(\omega_l-\omega_k)}$.

Let us now split the bound \eqref{eq:covariantbound}:
\begin{equation*}
\label{eq:covariantboundsplit}
|\rho'_{nm}| \leq \sideset{}{'}\sum_{\substack{c,d \\ \omega_c \leq \omega_n}}  |\rho_{cd}| \sqrt{p_{n|c} p_{m|d}} + \sideset{}{'}\sum_{\substack{c,d \\ \omega_c > \omega_n}}  |\rho_{cd}| \sqrt{p_{n|c} p_{m|d}}.
\end{equation*}
We can use the inequality \eqref{eq:transitionbound} in the first sum and use the time-translation symmetry condition \mbox{$\omega_c - \omega_d = \omega_n - \omega_m$}, that implies \mbox{$\omega_m - \omega_d=\omega_n - \omega_c \geq 0$}. Simple manipulations lead then to the final result given by: 
\be
\label{eq:thermalbound}
|\rho'_{nm}| \leq \sideset{}{'}\sum_{\substack{c,d \\ \omega_c \leq \omega_n}} |\rho_{cd}| e^{-\beta \hbar(\omega_{n}-\omega_c)}+\sideset{}{'}\sum_{\substack{c,d \\ \omega_c > \omega_n}} |\rho_{cd}|.
\ee
This bound on coherence tranformations by thermal operations can be easily interpreted physically. Time-translation symmetry implies that the contributions to $\rho'_{nm}$ can only come from elements within the same mode. The Gibbs-preserving condition (necessary for the non-existence of perpetuum mobiles) imposes an asymmetry in the contributions to the final coherence. The initial low-energy coherences, when contributing to the final high-energy coherences, are exponentially damped by the factor $e^{-\beta\hbar (\omega_n-\omega_c)}$. On the other hand our bound does not constrain the possibility of transforming high-energy coherences into coherences between lower energy levels. This irreversibility in coherence transformations can be best understood through elementary coherence manipulations, presented in the next section. Moreover, one can prove that the bounds presented in this section are tight for qubit systems and thus sufficient to fully solve the qubit interconversion problem under the restriction of either time-translation symmetric or thermal dynamics. We present these results together with the discussion of the temperature dependence of the set of achievable states in Appendix G.

\section{Applications to coherence transfer}
\label{sec:applications}

Previous works on coherence transformations under thermal maps \cite{cwiklinski2014limitations, narasimhachar2014low} have made the simplifying assumption that all energy differences in the Hamiltonian of the system are distinct. However, it is only the overall coherence in a mode (the sum of coherence terms) that has to decrease, not each off-diagonal term separately. Therefore, previous results do not capture all the physics of ubiquitous systems such as harmonic oscillators or spin-$j$ particles in a magnetic field, where modes are composed of more than one off-diagonal element. Our framework is suited to go beyond this restriction and reveals that within a given mode non-trivial dynamics takes place. However, thermodynamics imposes directionality on coherence transfers. 

To introduce these features it suffices to consider the simplest system with non-trivial mode structure -- a qutrit in a state $\rho$ described by the following Hamiltonian:
\begin{equation*}
H_S = \sum_{n=0}^2 n\hbar\omega_0 \ketbra{n}{n}.
\end{equation*}
Using Eq.~\eqref{eq:modes} we easily identify that the mode $\omega_0$ is composed of two off-diagonal elements: 
\begin{equation*}
\rho^{(\omega_0)} = \rho_{10} \ketbra{1}{0} + \rho_{21}\ketbra{2}{1},
\end{equation*}
while, e.g., \mbox{$\rho^{(2\omega_0)}=\rho_{20}\ketbra{2}{0}$} consists of a single term. We consider some primitive operations on this mode that may be used as building blocks in general coherence processing for higher-dimensional systems. One of them is \emph{coherence shifting}: shifting up or down in energy the coherence between two given energy levels, preserving as much of it as we can (e.g. $\rho_{10}$ can be shifted ``up'' to $\rho_{21}$, which can be then shifted ``down'' to $\rho_{10}$). Another primitive is \emph{coherence merging}: given two coherence terms (e.g. $\rho_{10}$ and $\rho_{21}$) one wants to optimally merge them into a single one (e.g. $\rho_{10}$). We will first study the limitations imposed by time-translation symmetry and then show how the situation changes in thermodynamics due to the second law.

\subsection{Coherence shifting under thermal operations}

Assume that the only non-vanishing coherence term is $|\rho_{10}|~=~c$ and that we want to transfer it inside mode $\omega_0$ to $|\rho_{21}|$, i.e., we want to transform the coherence between energy levels $\ket{0}$ and $\ket{1}$ into coherence between $\ket{1}$ and $\ket{2}$. Our bound \eqref{eq:covariantbound} for symmetric operations gives:
\be
\label{eq:symmetricshifting}
|\rho'_{21}| \leq c \sqrt{p_{1|0} p_{2|1}}\leq c.
\ee
If \eqref{eq:symmetricshifting} is tight, a perfect shift can be obtained. It is easy to check that this is actually the case: a symmetric map described by Kraus operators
\begin{subequations}
\begin{eqnarray}
M_1&=&\ketbra{1}{0}+\ketbra{2}{1},\label{eq:cov_shift1}\\
M_2&=&\ketbra{2}{2},\label{eq:cov_shift2}
\end{eqnarray}
\end{subequations}
perfectly shifts the coherence from $\ketbra{1}{0}$ to $\ketbra{2}{1}$. The situation would be analogous if we started with a coherence term $|\rho_{21}|$ and wanted to move it down in energy to $|\rho_{10}|$. Therefore, coherence transfer within a mode through symmetric operations is completely reversible.

\begin{figure}[t!]
\includegraphics[width=0.6\columnwidth]{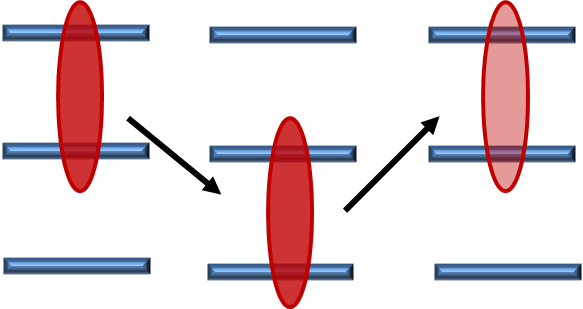}
\caption{\label{fig:shiftcycle}\emph{Irreversibility of coherence shift cycle.} Coherence between high energy levels is transferred down to low energy levels and then up again. The magnitude of the coherence terms is proportional to the intensity of the blobs. The first operation can be achieved perfectly, whereas the second results in damping of coherence. This directionality imposed by the second law implies that coherence transfers, similarly to heat transfers, are generally irreversible.}
\end{figure}

This reversibility breaks down in thermodynamics, where the second law requires Eq.~\eqref{eq:gp} to hold. We need to distinguish two situations: either we start with a coherence term $|\rho_{10}|=c$ and we move it up in energy to $|\rho_{21}|$ or we perform the reverse task. From Eq.~\eqref{eq:thermalbound} we immediately obtain a bound for the final magnitude of the transferred coherence:
\begin{subequations}
\begin{eqnarray}
\label{eq:shift1}
|\rho_{10}'|&\leq& c, \, \; \; \; \quad \quad \quad  \mathrm{for~shifting~down},\\
\label{eq:shift2}
|\rho_{21}'|&\leq& c e^{-\beta \hbar\omega_0},\quad \mathrm{for~shifting~up}.
\end{eqnarray}
\end{subequations}
Also in this case these bounds are tight, i.e., there are thermal operations achieving the above limits (see Appendix E). This proves that the irreversibility (directionality) within each mode suggested by Eq. \eqref{eq:thermalbound} is not just an artefact due to the bound being not tight. It is actually possible to perfectly transfer coherence down in energy, whereas the opposite task is exponentially damped due to the second law. Fig.~\ref{fig:shiftcycle} presents a ``shift cycle'', in which coherence between high energy levels is transferred down to lower energies and then up again. Due to the second law this thermodynamic process is irreversibile.

\begin{figure}[t!]
\includegraphics[width=0.6\columnwidth]{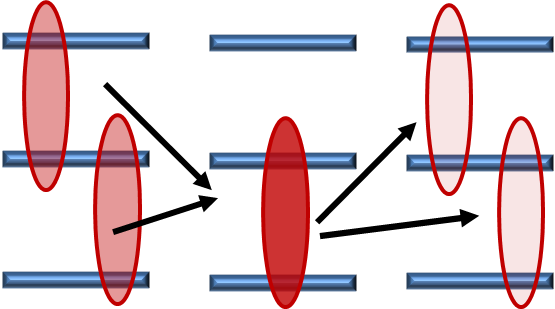}
\caption{\label{fig:merging}\emph{Irreversibility of coherence merging cycle.} Merging of coherences that are sharing an energy level always results in irreversible losses, even if we merge into the lower energy term. The second law, however, imposes additional irreversibility that exponentially damps the contribution to high energy coherence coming from low energy coherence.}
\end{figure}

\subsection{Coherence merging under thermal operations}

Let us now analyse a second primitive operation, coherence merging. Assume we are given a state $\rho$ with two non-vanishing coherence terms in mode $\omega_0$: \mbox{$|\rho_{10}|=a$} and \mbox{$|\rho_{21}|=b$}, and we want to merge them into a single coherence term $\rho_{10}'$ (the results for merging into $\rho'_{21}$ are analogous). Our bound \eqref{eq:covariantbound} for symmetric operations yields:
\begin{eqnarray*}
|\rho_{10}'|&\leq&\sqrt{p_{1|1}p_{0|0}}a+\sqrt{p_{1|2}p_{0|1}}b\leq\sqrt{p_{1|1}}a+\sqrt{p_{0|1}}b\nonumber\\
&\leq&\sqrt{p_{1|1}}a+\sqrt{1-p_{1|1}}b.
\end{eqnarray*}
One can easily prove that the above bound is maximized for $p_{1|1}=a^2/(a^2+b^2)$, so ultimately
\begin{equation}
\label{eq:merging_cov_bound2}
|\rho_{10}'|\leq \sqrt{a^2+b^2}.
\end{equation}
We note that a symmetric merging map achieving the above bound can actually be constructed (see Appendix F). It is also interesting to note that coherence merging at the maximum rate $a+b$ cannot be achieved (see Fig.~\ref{fig:merging}), as inevitable losses arise when the two coherence terms have an overlap, i.e., both correspond to the coherence between state $\ket{1}$ and one of the other two states. This property distinguishes merging from shifting. 

Let us now switch to the thermodynamic scenario. The bound for merging two coherences into a single coherence term now depends on whether one merges into high energy coherence or into low energy coherence. By applying a similar reasoning as in the case of symmetric operations we obtain bounds for coherence merging under thermal operations:
\begin{subequations}
\begin{eqnarray}
|\rho_{10}'|&\leq& \sqrt{a^2+b^2},\quad\quad\quad\; ~\mathrm{for~merging~down},\\
|\rho_{21}'|&\leq& \sqrt{e^{-\beta\hbar\omega_0}a^2+b^2},\; \; \mathrm{for~merging~up}.
\end{eqnarray}
\end{subequations}

Finally, let us note that the qutrit example does not exhaust all the merging scenarios. One of the reasons is that the non-trivial mode in the case analyzed is composed of two off-diagonal elements that are overlapping. For higher dimensional systems one can imagine a situation in which elements of the same mode are not overlapping, e.g., $\ketbra{1}{0}$ and $\ketbra{3}{2}$ for a system with equidistant spectrum. In contrast to the overlapping case for symmetric operations one can then perform perfect merging using the shift operation from the previous section, see Eqs.~\eqref{eq:cov_shift1} and \eqref{eq:cov_shift2}. However, we leave the comprehensive study of the set of building blocks for manipulating coherence for future research.

\section{Conclusions}

The present paper aimed at several things. The broad approach was to analyse coherence manipulations in thermodynamics from a symmetry-based perspective. Specifically, the underlying energy-conservation within thermodynamics was shown to constrain all thermodynamic evolutions to be ``symmetric'' under time-translations in a precise sense. This in turn allowed us to make use of harmonic analysis techniques, developed in \cite{marvian2013modes}, to track the evolution of coherence under thermodynamic transformations in terms of the ``mode components'' of the system. This constitutes a natural framework to understand coherence, thus allowing us to separate out the constraints that stem solely from symmetry arguments from those particular to thermodynamics, and provides results that generalize recent work on coherence \cite{cwiklinski2014limitations}. This approach also implies that the existing single-shot results applicable to block-diagonal results, constrained by thermo-majorization, can be viewed as particular cases of our analysis when only the zero-mode is present. Beyond this regime we have shown that every non-zero mode obeys \emph{independent} constraints and displays thermodynamic irreversibility similar to the zero-mode.

Exploiting these tools we arrived at inequalities linking initial and final coherences in the energy eigenbasis. We have shown that a rich dynamics is allowed, in which coherence can be transferred among different energy levels within each mode, and that, similarly to heat flows, coherence flows show directionality due to the limitations imposed by the second law. This new kind of irreversibility adds up to the ones identified in work extraction \cite{horodecki2013fundamental} and coherence distillation \cite{lostaglio2015description}. Finally, we have also presented a way to find the guaranteed amount of coherence that can always be preserved under thermodynamic transformations. 

\bigskip

\textbf{Acknowledgements:} ML and KK would like to thank Mercedes Gimeno-Segovia and Oscar Dahlsten for helpful discussions. KK is supported by EPSCR and in part by COST Action MP1209, ML is supported in part by EPSCR, COST Action MP1209 and Fondazione Angelo Della Riccia. DJ is supported by the Royal Society. TR is supported by the Leverhulme Trust.

\bibliography{Bibliography_thermodynamics}

\clearpage

\section*{Appendix}

We provide various results and elaborations that are relevant to the main text.

\subsection{Interpretations of the first law of thermodynamics}

One may argue that our analysis relies on the interpretation of the first law of thermodynamics. Indeed, there are at least two ways of formalizing it: as strong energy conservation (the total Hamiltonian of system and environment is a conserved charge, e.g., \cite{janzing2000thermodynamic, horodecki2013fundamental, skrzypczyk2013extracting} and the present work) and weak energy conservation (the average energy is conserved only for a given initial state \cite{skrzypczyk2014work}). Weak energy conservation allows a larger class of free thermodynamic operations, however this comes at the price: the allowed operations are now \emph{state-dependent}, which is theoretically undesirable. Moreover, this class of operations assumes an unbounded availability of coherence, which may or may not be freely accessible in extreme quantum regimes. Recently, \AA{}berg has shown that work can be catalytically released from quantum coherence under strict energy conservation, provided that we have a sufficiently large coherence resource \cite{aberg2014catalytic}. This suggests that the results obtained under the assumption of weak energy conservation can be recovered in the framework of strict energy conservation, with the advantage of taking explicitly into account the coherent resources used.

\subsection{Arbitrary free energy trivializes thermal operations to symmetric operations}

Consider thermal operations, as defined in the main text. Suppose now we allow arbitrary copies of pure, zero-coherence states. The assumption of zero coherence means that the resulting theory is a subset of time-translation symmetric operations. Conversely, any symmetric operation $\E$ possesses a Stinespring dilation \mbox{$\E (X ) = \tr{E}{U ( X \otimes \sigma_E )U^\dagger}$} where $\sigma_E$ is a symmetric state and $U$ is a symmetric unitary on the joint system. Thermal operations on $|0\rangle^{\otimes N}$ (for arbitrary $N$) allow for the creation of any symmetric state, and so $\sigma_E$ can be formed. Therefore the above evolution can be realised and so the theory is trivialised to the theory of time-translation symmetric operations when an arbitrary amount of work is provided.

\subsection{Thermal operations form a convex set}

The proof is as follows. Let $\E_1$ and $\E_2$ be two thermal maps acting on a system $S$ defined as in Eq.~\eqref{eq:thermal}. $\mathcal{E}_1$ is defined by $(U_1, \gamma_1)$ and $\mathcal{E}_2$ by $(U_2, \gamma_2)$ where
\be
\gamma_1 = \frac{e^{-\beta H_{1}}}{Z_1}, \quad \gamma_2 = \frac{e^{-\beta H_{2}}}{Z_2}
\ee
and $U_i$ is an energy preserving unitary on $S+E_i$:
\be
[U_i, H_S + H_{i}]= 0, \quad i=1,2.
\ee
We now show that a linear combination
\be
p \mathcal{E}_1 + (1-p)\mathcal{E}_2
\ee
is a thermal operation. Let us introduce a $d$-dimensional ancillary bath state $\gamma_A$ with Hamiltonian $H_A=\I_d$ and a joint unitary acting on system, the two environments and the ancilla, $S+ E_{1} + E_{2} +A$. The total Hamiltonian of this joint system is $H=H_S + H_{1} + H_{2} + H_A$. Now define the controlled unitary
\begin{equation}
U := \Pi_1 \otimes U_1 + \Pi_2 \otimes U_2,
\end{equation}
where $\Pi_1$ and $\Pi_2$ are respectively rank $k$, and rank $d-k$ projectors onto the degenerate bath system of the ancilla $A$ and $\Pi_1 + \Pi_2 = \I_d$. We can check that, for $i=1,2$,
\begin{eqnarray}
\nonumber
[\Pi_i \otimes U_i, H]  &= & [\Pi_i \otimes U_i, H_S + H_{i}] \\
\nonumber
& = & \Pi_i \otimes [U_i, H_S + H_{i}] =0,
\end{eqnarray}
so that $U$ is energy-preserving on $S+ E_{1} + E_{2} + A$. We finally have
\begin{eqnarray*}
& & \tr{A,E_1,E_2}{U ( \rho \otimes \gamma_A \otimes \gamma_1 \otimes \gamma_2)U^\dagger} \\
&= &\frac{1}{d}\sum_{i=1}^2 \tr{A,E_i}{\Pi_i \otimes U_i ( \rho \otimes \I_d \otimes \gamma_i)\Pi_i \otimes U^{\dagger}_i}  \\
&= & \frac{k}{d} \E_1 (\rho) + \left(1-\frac{k}{d}\right) \E_2 (\rho).
\end{eqnarray*}
Thus $(U, \gamma_A \otimes \gamma_1 \otimes \gamma_2)$ defines a thermal operation equivalent to any rational convex combination of $(U,\gamma_1)$ and $(U_2, \gamma_2)$. Irrational convex combinations are approached with arbitrary accuracy.

\subsection{Gibbs-preserving condition}

Here we prove Eq.~\eqref{eq:transitionbound} here. From Eq.~\eqref{eq:GP_cond} after simple transformations one obtains that for every $l$
\begin{equation*}
p_{l|l}=1-\sum_{i\neq l} p_{l|i} \frac{r_i}{r_l}.
\end{equation*}
Taking into account that $p_{l|l}$ is positive (as it represents transition probability) yields for every $k\neq l$
\begin{equation*}
p_{l|k}\leq \frac{r_l}{r_k}-\sum_{i\neq l,k} p_{l|i} \frac{r_i}{r_k}\leq\frac{r_l}{r_k}=e^{\beta\hbar(\omega_k-\omega_l)}.
\end{equation*}

\subsection{Coherence shifting by thermal operations}

Here we present how to construct thermal operations that achieve the bounds \eqref{eq:shift1} and \eqref{eq:shift2} for shifting the coherence. In both cases (moving the coherence term up and down in energy) we can use a bath state given by
\begin{equation*}
\gamma=\frac{1}{Z}\sum_{n=0}^{\infty}e^{-\beta n \hbar\omega}\ketbra{n}{n},
\end{equation*}
with partition function $Z=(1-e^{-\beta \hbar\omega})^{-1}$. Now, consider the following joint unitary:
\small
\begin{eqnarray*}
U&=&\ketbra{00}{00}+\ketbra{01}{10}+\ketbra{10}{01}\\
&+&\sum_{i=2}^{\infty}\ketbra{1;i-1}{2;i-2}+\ketbra{0;i}{1;i-1}+\ketbra{2;i-2}{0;i}.
\end{eqnarray*}
\normalsize
It is easy to see that the above unitary is energy conserving, as it only mixes states with the same total energy. By direct calculation we can now check that
\begin{eqnarray*}
\tr{E}{U(\ketbra{2}{1}\otimes\gamma)U^{\dagger}}&=&\ketbra{1}{0},\\
\tr{E}{U^{\dagger}(\ketbra{1}{0}\otimes\gamma)U}&=&e^{-\beta\hbar\omega}\ketbra{2}{1}.
\end{eqnarray*}
Hence both bounds, for shifting down in energy (Eq.~\eqref{eq:shift1}) and up in energy (Eq.~\eqref{eq:shift2}), are achievable via the presented thermal operations.

\subsection{Coherence merging by symmetric operations}

Here we will construct a symmetric operation achieving the bound \eqref{eq:merging_cov_bound2} for merging coherence. Consider the following Kraus operator decomposition of a CPTP map:
\begin{eqnarray*}
M_j&=&\frac{1}{\sqrt{3}}\left[ \ket{0}\left(e^{i\frac{2\pi j}{3}}\bra{0}+x\bra{1}\right)\right]\\
&+&\left.\ket{1}\left(e^{i\frac{2\pi j}{3}}\sqrt{1-x^2}\bra{1}+\bra{2}\right)\right],\\
\end{eqnarray*}
with $x\in[0,1]$ and $j=\{0,1,2\}$. It is easy to show that this map is time-translation symmetric by checking that each mode is mapped into itself. By direct calculation we can now show that
\begin{eqnarray*}
|\rho_{10}'|&=&\bra{1}\left(\sum_{j=0}^{2}M_j(a\ketbra{1}{0}+b\ketbra{2}{1})M_j^{\dagger}\right)\ket{0}\\
&=&\sqrt{1-x^2}a+xb.
\end{eqnarray*}
The choice $x=b/\sqrt{a^2+b^2}$ saturates the bound \eqref{eq:merging_cov_bound2}.

\subsection{State interconversion limitations for qubit systems}
\label{sec:qubit}

The requirement for available maps to respect the laws of thermodynamics constrains the allowed dynamics, so that not all state transformations are possible. For example, we have shown in Sec. \ref{sec:applications} that thermal maps allow us to shift coherence up in energy only at the price of an exponential damping. A general question that one can ask in the scenario of constrained dynamics is the interconversion problem: given an initial state $\rho$ what is the set of states $\{\sigma\}$ achievable via the allowed maps? Here we analyse this for a qubit system contrasting symmetric and thermal transformations. In the latter case we also highlight the dependence of coherence preservation on the temperature of the bath.

Let us first parametrize the initial state of the qubit system $\rho$ and its final state $\sigma$, written in the eigenbasis of the Hamiltonian, in the following way
\begin{equation*}
\rho=\left(\begin{array}{cc}
p&c\\
c&1-p
\end{array}\right),\quad
\sigma=\left(\begin{array}{cc}
q&d\\
d&1-q
\end{array}\right),
\end{equation*}
where $c$ and $d$ are assumed real without loss of generality, as a phase change in the coherence terms is both symmetric and conserves energy. The bound \eqref{eq:covariantbound} for symmetric operations yields
\begin{equation}
\label{eq:qubit_bound}
d\leq c\sqrt{p_{0|0}p_{1|1}}.
\end{equation}
To obtain a distribution $\v{q}=(q,1-q)$ from $\v{p}=(p,1-p)$ the transition matrix $\Lambda$, defined by transition probabilities $p_{j|i}$ with $i,j\in\{0,1\}$, must fulfill $\Lambda\v{p}=\v{q}$. This condition together with the stochasticity of $\Lambda$ gives
\begin{subequations}
\begin{eqnarray*}
p_{0|0}&=&\frac{(p_{1|1}-1)(1-p)+q}{p}\leq\frac{q}{p},\\
p_{1|1}&=&\frac{(p_{0|0}-1)p+1-q}{1-p}\leq\frac{1-q}{1-p}.
\end{eqnarray*}
\end{subequations}
Note that for $q<p$ only the first inequality is nontrivial, whereas for $q>p$ only the second inequality is nontrivial. Using these conditions in Eq.~\eqref{eq:qubit_bound} gives:
\begin{figure}[t!]
\includegraphics[width=\columnwidth]{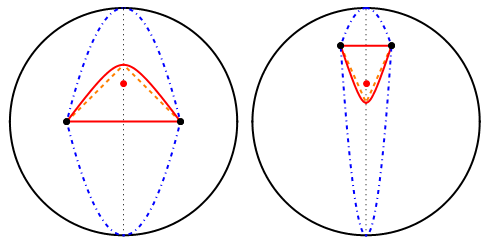}
\caption{\label{fig:qubit_cov_th} Extremal achievable states from a given initial state $\rho$ (black points) under time-translation invariant (blue dot-dashed lines) and thermal (solid red lines) operations presented on a Bloch sphere. Dotted lines join the eigenstates of system Hamiltonian and the red points correspond to thermal occupation of the ground state, here chosen to be $r=2/3$. Dashed orange lines correspond to the set of states obtained using the bound for guaranteed coherence preservation.}
\end{figure}
\begin{figure}[t!]
\includegraphics[width=\columnwidth]{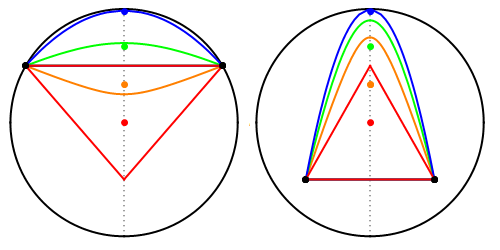}
\caption{\label{fig:qubit_thermal} Extremal achievable states from a given initial state $\rho$ (black points) under thermal operations at different temperatures (colorful lines) presented on a Bloch sphere. Dotted lines join the eigenstates of system Hamiltonian and different colors correspond to different thermal occupations. The points on the $z$ axis represent thermal states for the same set of temperatures (the red point in the centre corresponds to infinite-temperature bath, whereas the blue one at the boundary corresponds to the low-temperature limit bath).}
\end{figure}
\begin{equation}
\label{eq:qubitasymmetry}
d\leq c\sqrt{\alpha},
\end{equation}
where \mbox{$\alpha=\min\left(\frac{q}{p},\frac{1-q}{1-p}\right)$}. One can check that the time-translation symmetric CPTP map given by the following Kraus operators:
\begin{eqnarray*}
M_1&=&\ketbra{0}{0}+\sqrt{\alpha}\ketbra{1}{1},\\
M_2&=&\sqrt{1-\alpha}\ketbra{0}{1},
\end{eqnarray*}
saturates this bound for $q>p$, whereas a CPTP map given by $\{XM_1X,XM_2X\}$, with $X=\ketbra{0}{1}+\mathrm{h.c.}$, saturates the bound for $q<p$. Of course if we can saturate the bound, we can also obtain all states with coherence smaller than maximal, simply by partially dephasing the optimal final state (partial dephasing is a symmetric operation). This shows that the bound of Eq.~\eqref{eq:qubitasymmetry} captures all the constraints imposed by time-translation symmetry on the evolution of qubit states (a question left open in \cite{marvian2014extending}). In Fig.~\ref{fig:qubit_cov_th} we depict the extremal set of obtainable states via symmetric dynamics on a Bloch sphere for exemplary initial states (blue dot-dashed lines). 

We will now focus on thermal maps and see how the condition $\Lambda\v{r}=\v{r}$, changes the picture in thermodynamics ($\v{r}$ stands here, as in the main text, for the vector of thermal occupation probabilities). The choice of $\v{p}$ and $\v{q}$, together with the Gibbs-preserving condition, completely fixes $\Lambda$:
\begin{eqnarray*}
p_{0|0}&=&\frac{q(1-r)-r(1-p)}{p-r},\\
p_{1|1}&=&\frac{r(1-q)-p(1-r)}{r-p}.
\end{eqnarray*}
Hence, from Eq.~\eqref{eq:qubit_bound} we obtain
\begin{equation}
d\leq c\frac{\sqrt{(q(1-r)-r(1-p))(p(1-r)-r(1-q))}}{|p-r|}.
\end{equation}
The above bound has been recently shown to be tight \cite{cwiklinski2014limitations}, i.e., there exists a thermal operation that saturates it. In Fig.~\ref{fig:qubit_cov_th} we depict the extremal set of obtainable states via thermal operations on the Bloch sphere for exemplary initial states (red solid lines). 

Let us now proceed to the guaranteed coherence bound. Using the thermomajorization condition for a qubit we find that the extremal achievable incoherent states, characterized by probability distribution \mbox{$\tilde{\v{q}}=(\tilde{q},1-\tilde{q})$}, are given by 
\begin{eqnarray*}
\tilde{q}&=&1-\frac{1-r}{r}p,\quad \;\;\,\mathrm{for~}p>r,\\
\tilde{q}&=&\frac{r}{1-r}(1-p),\quad \mathrm{for~}p<r.
\end{eqnarray*}
Then $\lambda_*$, specified in Eq. \eqref{eq:guaranteed}, is given by
\begin{equation*}
\lambda_*=\frac{q-\tilde{q}}{p-\tilde{q}},
\end{equation*}
so that it is always possible to preserve at least $d=\lambda_*c$ coherence, while thermodynamically transforming the probability distribution over the energies from $\v{p}$ to $\v{q}$. The set of states obtained using the bound for guaranteed coherence preservation is depicted in Fig. \ref{fig:qubit_cov_th} (orange dashed lines). 

In Fig.~\ref{fig:qubit_thermal} we compare the set of obtainable states for different thermal distributions, i.e., for different temperatures. We make two interesting observations concerning thermal dependence of coherence preservation. Firstly, note that as $r$ approaches 1 (the temperature goes to zero, which is the limit recently studied in \cite{narasimhachar2014low}) the set of states obtainable via thermal operations coincides with ``half'' of the set of states obtainable via time-translation symmetric operations - as long as $q>p$ one can preserve the same amount of coherence. This suggests that the limitations of low-temperature thermodynamics can be inferred from the limitations on symmetric operations that are studied by the resource theory of asymmetry. Secondly, let us distinguish between heating processes (when $q<p$) and cooling processes (when $q>p$). Then one can check that in the heating scenario the higher the temperature of the bath, the more coherence one can preserve, whereas for cooling processes, the lower temperature ensures better coherence preservation. This shows that for general thermodynamic state transformations to optimally preserve coherence it is necessary to use baths of different temperatures.

\end{document}